# ResNet-34 with Lightweight Decoder for Accurate and Efficient Segmentation of Fetal Brain MRI

Ashiqur Rahman[1,3], Muhammad E. H. Chowdhury[2, *], Md. Abu Sayed[1], Md. Sharjis Ibne Wadud[1], Abu Naser Md. Arafat[3], Mehedi Hasan Prince[3]

[1]Department of Biomedical Physics and Technology, University of Dhaka, Dhaka, Bangladesh. Email: ashiq@bmpt.du.ac.bd (AR); sharjis@du.ac.bd (MSIW); sayed.kuet.bme@gmail.com (MAS);

[2]Department of Electrical Engineering, College of Engineering, Qatar University, Doha, Qatar. Email: mchowdhury@qu.edu.qa;

[3]Department of Biomedical Engineering, Jashore University of Science and Technology, Jashore, Bangladesh. Email: naser05bmearafat@gmail.com (ANMA); mhprince.bme@gmail.com (MHP);

[*]Corresponding author: mchowdhury@qu.edu.qa

## Abstract

Accurate segmentation of fetal brain tissues in Magnetic Resonance Imaging (MRI) is critical for early diagnosis of congenital abnormalities and improving prenatal care. However, the task remains difficult because of fetal motion, low tissue contrast, and major anatomical variability throughout gestational ages, particularly in segmenting complex structures such as white matter, gray matter, lateral ventricles, deep gray matter, extra-cerebrospinal fluid, cerebellum, and brainstem. As a solution to these difficulties, this research introduces a novel deep learning model that combines a ResNet-34 encoder with a lightweight decoder leveraging multi-layer perceptron (MLP) modules for adaptive feature refinement. This design specifically enhances the model's ability to preserve anatomical boundaries and mitigate segmentation errors caused by motion artifacts and intensity inhomogeneities. Computational efficiency is achieved by reducing parameter count, employing bilinear upsampling instead of transposed convolutions, and optimizing the decoder for speed without sacrificing accuracy. Trained and validated on the FeTA 2021 dataset using 5-fold cross-validation, the proposed model outperforms baseline architectures such as UNet, UNet++, DeepLabV3, and DeepLabV3+, achieving an average Accuracy of 97.37% with a mean Dice Similarity Coefficient (DSC) of 90.33%, mean Intersection over Union (IoU) of 86.93%, and Precision of 90.83%. Additionally, its fast inference time and reduced computational load make it well-suited for integration into real-time clinical workflows.

## 1. Introduction

Monitoring and understanding fetal brain development is essential for detecting early signs of neurodevelopmental disorders and improving pregnancy outcomes. During gestation, the fetal brain experiences quick and complex morphological modifications, and disruptions during this period can have long-term consequences on cognitive, motor, and behavioral functions (Edde et al., 2021; Mahin et al., 2024a; Sathyanesan et al., 2019; Ten Donkelaar et al., 2014). Magnetic Resonance Imaging (MRI) has surfaced as the modality of choice for fetal brain evaluation for high-resolution, non-invasive capabilities and excellent soft-tissue contrast (Di Mascio et al., 2022; Papaioannou et al., 2021). Proper fetal brain tissue segmentation in MRI is critical for detecting congenital abnormalities and assessing developmental trajectories. Automated segmentation techniques can significantly enhance diagnostic efficiency and consistency.

Manual annotation of fetal MRI scans is time-consuming, labor-intensive, and often prone to irregularity due to subjective interpretation. Even among expert radiologists, annotations may vary significantly when faced with low-quality or motion-degraded images. The complexity of fetal-brain morphology, combined with differences in gestational age, further exacerbates these inconsistencies, leading to challenges in reproducibility and diagnostic reliability (Hutter et al., 2019; Mahin et al., 2024b; Shen et al., 2022; Uus et al., 2023). To solve these problems, deep learning methods have been increasingly applied for automated segmentation in fetal brain imaging. The UNet architecture (Ronneberger et al., 2015) has laid the groundwork for medical image segmentation and remains a dominant baseline because of its skip-connection encoder-decoder structure. UNet++ (Zhou et al., 2018) enhanced this approach with nested dense skip connections to improve feature fusion. DeepLabV3 (L.-C. Chen et al., 2017) and DeepLabV3+ (L. C. Chen et al., 2018) introduced atrous spatial pyramid pooling and encoder-decoder structures to effectively capture contextual information at multiple scales. Transformer-based models have become more popular in recent years for medical segmentation. CNNs and transformers are combined in TransUNet (J. Chen et al., 2021a) to capture not only local but also global features to enhance segmentation in complex structures. Swin-UNet (Cao et al., 2021) leverages shifted window transformers to enhance hierarchical feature representation. SegFormer (Xie et al., 2021) simplifies segmentation architecture by removing positional encoding while maintaining competitive accuracy and efficiency. These advances have influenced recent fetal brain segmentation studies. A variety of advanced architectures have been employed in recent literature, such as contextual transformer-based models for multiclass segmentation (X. Huang et al., 2023), real-time 2D UNet frameworks to handle motion artifacts (Salehi et al., 2018), multiscale feature fusion with graph convolution mechanisms (Qi et al., 2024), and optimized 3D U-Net models using section-to-volume reconstruction (Zhao et al., 2022). Other efforts include multi-view inception residual networks combined with radiomic features (Mazher et al., 2022), UNet and Attention UNet architectures across multiple MRI modalities (Faghihpirayesh et al., 2024), refined nnUNet configurations (Peng et al., 2023), hybrid FCNs integrating multi-channel inputs with rigid alignment (Li et al., 2020), GAN-based models with 3D U-Net backbones for fMRI (De Asis-Cruz et al., 2022), and a ResUNet-based approach was proposed for fetal cerebellum segmentation in 2D ultrasound, achieving high precision but without extension to multi-class segmentation in MRI (V. Singh et al., 2021).

Though deep learning has advanced, fetal brain MRI remains challenging because of biological and technical factors. Low tissue contrast between developing brain structures, spontaneous fetal motion during image acquisition, and anatomical variability across gestation contributes to poor image quality and inconsistent segmentation performance. These factors hinder the generalizability and reliability of existing models, particularly in clinical settings where motion artifacts and imaging protocol differences are common. Consequently, there remains a critical need for segmentation methods that preserve anatomical accuracy under challenging conditions while maintaining computational efficiency for real-time deployment in clinical workflows.

To address these challenges, this study proposes a novel fetal brain tissue segmentation framework that integrates a ResNet-34 encoder with an MLP-based decoder. ResNet-34 is selected for its balanced trade-off between representational capacity and computational efficiency, offering sufficient depth for hierarchical feature extraction while avoiding the overfitting risk and memory overhead associated with deeper variants such as ResNet-50 or ResNet-101. This design choice is particularly suitable for fetal MRI applications characterized by limited annotated data and practical clinical constraints. The objective of this work is to develop an efficient fetal brain tissue segmentation architecture that combines robust hierarchical feature extraction with refined decoding, while improving segmentation performance under challenging imaging conditions such as motion artifacts and low tissue contrast. In addition, the study aims to achieve computational efficiency that supports time-sensitive clinical analysis without sacrificing segmentation accuracy or anatomical fidelity.

The main contributions of this study are summarized as follows:

- A novel deep learning architecture that integrates a ResNet-34 encoder for hierarchical feature extraction with an MLP-based decoder for refined fetal brain tissue segmentation.
- Improved segmentation accuracy through adaptive feature fusion, effectively reducing errors caused by motion artifacts and low tissue contrast.
- A computationally efficient design that supports practical deployment in clinical fetal brain analysis.

The remainder of this paper is organized as follows. Section 2 reviews related work on fetal brain segmentation, highlighting recent advances and remaining limitations. Section 3 presents the proposed methodology, including preprocessing steps, dataset details, and model architecture. Section 4 presents the experimental results, including comparisons with baseline models, computational complexity analysis, sensitivity analysis under varying contrast and motion conditions, and ablation studies. Finally, Section 5 concludes the paper and discusses future research directions.

## 2. Related Work

Automated fetal brain MRI segmentation has received increasing attention due to its critical role in assessing neurodevelopment and supporting prenatal diagnosis. Early approaches relied on atlas-based and registration-driven methods, which were limited by sensitivity to fetal motion, anatomical variability across gestational ages, and computational complexity. With the advent of deep learning, convolutional neural network-based architectures, particularly U-Net and its variants, have become the dominant paradigm for fetal brain segmentation tasks due to their ability to learn hierarchical representations from data (Ronneberger et al., 2015; Zhou et al., 2018).

Despite notable progress, existing fetal brain segmentation methods continue to face several qualitative limitations that restrict their clinical applicability. One major challenge is motion artifacts, which are inherent to in utero MRI acquisition and can severely degrade image quality, leading to blurred boundaries and incomplete tissue representation (Keraudren et al., 2014; Mohseni Salehi et al., 2017) Additionally, low tissue contrast, particularly at interfaces such as gray matter and extra-cerebrospinal fluid or gray and white matter, complicates accurate boundary delineation, even for advanced architectures (Huang et al., 2023). These issues are further exacerbated by anatomical variability across gestational ages, where rapid morphological changes alter tissue appearance and spatial relationships (Ten Donkelaar et al., 2014).

Another important limitation is domain shift across scanners, reconstruction pipelines, and imaging protocols, which can significantly affect model generalization. Models trained on narrowly scoped datasets often exhibit degraded performance when applied to data from different acquisition settings or reconstruction strategies (Faghihpirayesh et al., 2022; S. Huang et al., 2023). Moreover, fetal brain segmentation datasets, including FeTA 2021, are characterized by annotation uncertainty and inter-rater variability, particularly for small or low-contrast structures such as ventricles and deep gray matter. This label ambiguity introduces noise during training and challenges the stability of complex or highly parameterized models (Payette et al., 2021).Recent studies have explored increasingly sophisticated architectures, including transformer-based and hybrid CNN–transformer models, to capture long-range dependencies and global context (Cao et al., 2021; Chen et al., 2021). While these methods show promise, they often require larger training datasets and incur substantial computational cost, limiting their practicality in fetal MRI settings where annotated data is scarce and real-time deployment is desirable. Similarly, fully 3D convolutional approaches can exploit volumetric context but suffer from high memory demands, reduced effective sample size, and increased sensitivity to reconstruction artifacts under limited data conditions (S. P. Singh et al., 2020).

Taken together, these limitations highlight a clear research gap. Existing methods often prioritize architectural complexity or marginal performance gains without adequately addressing robustness to motion artifacts, annotation uncertainty, and computational efficiency under realistic clinical constraints. There remains a need for a segmentation framework that balances anatomical fidelity, stability under limited and heterogeneous data, and efficiency suitable for practical deployment. Motivated by these challenges, the present work proposes a lightweight yet expressive segmentation architecture that integrates a ResNet-based encoder with an efficient decoder design, aiming to improve segmentation robustness and accuracy while maintaining computational practicality for fetal brain MRI analysis.

## 3. Methodology

This section presents a summary of the systematic approach employed in this study to achieve our objectives. It gives a thorough explanation of the dataset, pre-processing steps, model architecture, training procedures, and evaluation metrics. The workflow aims to ensure robustness, reproducibility, and comparability of results, leveraging state-of-the-art techniques in fetal brain tissue segmentation. The following sections elaborate on each stage of the methodology, supported by a visual representation of the workflow in Figure 3.1.

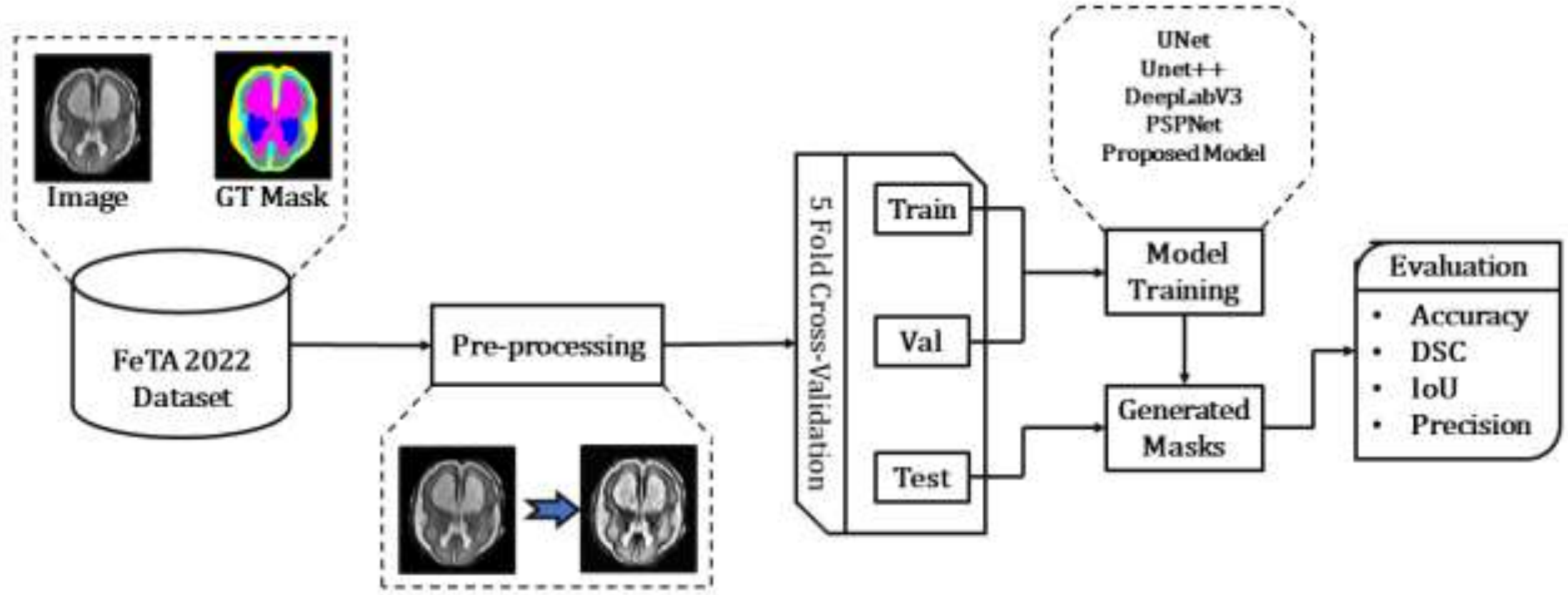

Figure 3.1: Schematic overview of the fetal brain tissue segmentation pipeline. The diagram outlines the key steps, from data preparation and model training to evaluation.

### 3.1 Dataset Description

This study employs the Fetal Tissue Annotation (FeTA) 2021 dataset, a publicly available, expertly curated resource designed to support the creation of automated multi-tissue segmentation algorithms for MRI of the fetal brain (Payette et al., 2021). The dataset comprises 50 super-resolved 3D fetal brain volumes reconstructed from T2-weighted single-shot fast spin echo (ssFSE) sequences, covering a gestational age range of 20 to 33 weeks. It includes both pathological (e.g., spina bifida) and non-pathological cases, offering a representative sample of clinical variability. Each volume contains 256 high-resolution axial slices with manual annotations into seven critical brain tissue classes: external cerebrospinal fluid, brainstem/spinal cord, cerebellum, grey matter, white matter, deep grey matter, and ventricles. The dataset is divided into 10 unlabeled volumes for testing and 40 labeled volumes for training. These detailed labels are created through expert-guided manual segmentation and post-processed using interpolation and quality review, facilitate fine-grained multiclass segmentation tasks.

FeTA 2021 was selected due to its combination of anatomical diversity and rich expert annotations, which collectively enable both robust model training and rigorous benchmarking. Its inclusion of both pathological and non-pathological cases across a broad gestational age range enhances the potential for generalizable model development. However, the dataset presents certain limitations. First, the total sample size is relatively modest, which may limit statistical generalization. Second, gestational age distribution is uneven, with underrepresentation at the extremes (e.g., ≤20 weeks or ≥34 weeks). Third, only a small subset of volumes underwent multi-rater segmentation, limiting analysis of inter-rater variability. Lastly, while motion artifacts are present, they are not uniformly distributed, potentially biasing model performance toward higher-quality cases. Despite these constraints, FeTA 2021 remains one of the most comprehensive and clinically relevant datasets available for fetal brain segmentation, serving as a robust foundation for training and validating models aimed at real-world prenatal diagnostic applications.

In Figure 3.2, three samples axial view of a fetal brain MRI scan is shown alongside its corresponding annotated mask. The figure illustrates the precision of the manual segmentations, with distinct labels for each tissue class, including grey matter, white matter, lateral ventricles, brainstem, deep grey matter, cerebellum, and external cerebrospinal fluid.

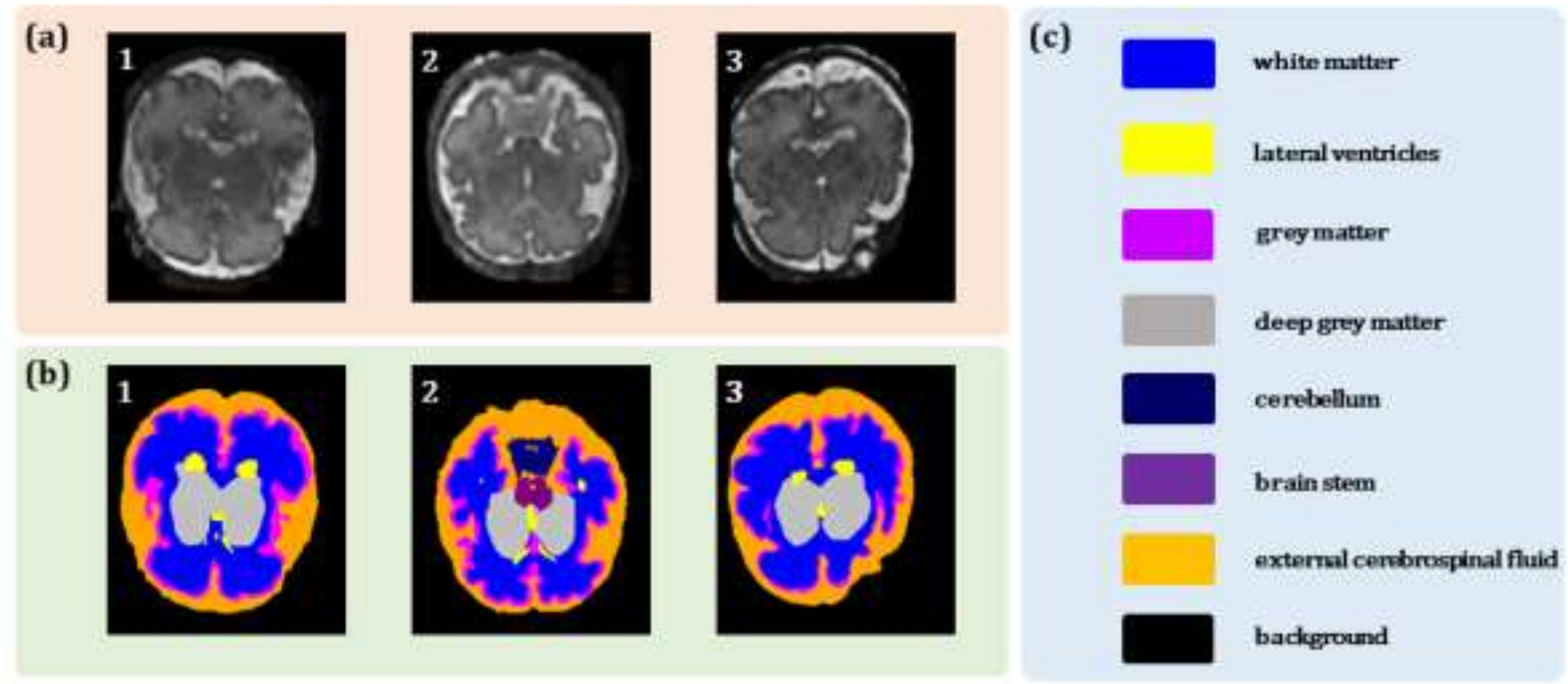

Figure 3.2: Axial view of a fetal brain MRI scan and its corresponding annotated mask. (a) Original fetal brain MRI scan. (b) Annotated mask showing the segmentation of seven tissue classes: white matter, grey matter, lateral ventricles, deep grey

matter, cerebellum, brainstem, and external cerebrospinal fluid. (c) Color-coded legend for the annotated tissue classes with background.

### 3.2 Data Preprocessing

The preprocessing pipeline was designed to standardize the fetal brain MRI data and enhance image quality for subsequent analysis. First, the original MRI volumes, stored in Neuroimaging Informatics Technology Initiative (NIfTI) format, were changed into 2D axial slices in Portable Network Graphics (PNG) format using Python libraries, nibabel for NIfTI file handling and OpenCV for image resizing. Each slice was resized to a resolution of 256×256 pixels for dataset uniformity. Next, Contrast Limited Adaptive Histogram Equalization (CLAHE) was a to enhance tissue boundaries and improve image contrast (Rahman et al., 2025). CLAHE improves image contrast by dividing it into small regions and enhancing each one separately (Kuran & Kuran, 2021; Sami et al., 2025). This prevents over-amplification of noise while enhancing local contrast.

For each tile, the histogram $\mathrm{H(i)}$ is computed, where i represents the intensity levels. The histogram is clipped to a predefined limit L to avoid excessive amplification of noise. The clipped histogram is then equalized using the cumulative distribution function (CDF):

$$\mathrm{CDF}\ (\mathrm{i}) = \sum_{\mathrm{j}=0}^{\mathrm{i}} \mathrm{H(i)} \quad (1)$$

The equalized intensity values are mapped to the output range, and bilinear interpolation is applied to smooth the boundaries between tiles. Mathematically, the output intensity $\mathrm{I_{out}(x, y)}$ at pixel $\mathrm{(x, y)}$ is given by:

$$\mathrm{I_{out}(x, y)} = \mathrm{round}\ \left(\frac{\mathrm{CDF}\big(\mathrm{I_{in}(x, y)}\big) - \mathrm{CDF_{min}}}{\mathrm{M \times N - CDF_{min}}}\right) \times (\mathrm{L} - 1) \quad (2)$$

Where $\mathrm{I_{in}(x, y)}$ is the input intensity, $\mathrm{M \times N}$ is the total number of pixels in the tile, and L is the number of intensity level? CLAHE was implemented with a clip limit of 2.0 and a tile grid size of 8×8. This step significantly improved the visibility of fine structures in the fetal brain MRI images, facilitating more accurate segmentation.

However, CLAHE can introduce certain drawbacks. Over-enhancement in uniform regions may lead to artificial textures, and aggressive parameter settings can exaggerate noise or produce unnatural boundaries. To mitigate these risks, parameters were optimized through iterative testing on various gestational age ranges and tissue contrasts.

Additionally, class-wise binary masks were merged into a single multiclass segmentation mask by assigning fixed integer labels to each anatomical structure based on a predefined mapping. For each pixel, label assignment was determined by the presence of non-zero values in the respective class mask. The combined masks were saved as 8-bit PNG images with a standardized color palette, ensuring consistency and compatibility for model training and evaluation.

## 3.3 Proposed Model Architecture

The proposed model architecture for fetal brain tissue segmentation is illustrated in Figure 3.3. The architecture combines a ResNet-34 encoder with a MLP based decoder designed to efficiently fuse multi-scale features and produce high-resolution segmentation maps. Hierarchical features are extracted by the encoder from the input MRI images using residual blocks, while the decoder employs MLP modules and bilinear interpolation to refine and upsample the feature maps. The figure provides a detailed visual representation of the components of the model, including the residual blocks in the encoder, the MLP modules in the decoder, and the feature fusion process. This architecture is specifically designed to handle the complex and rapidly changing anatomy of the developing fetal brain, ensuring precise localization and classification of tissue boundaries.

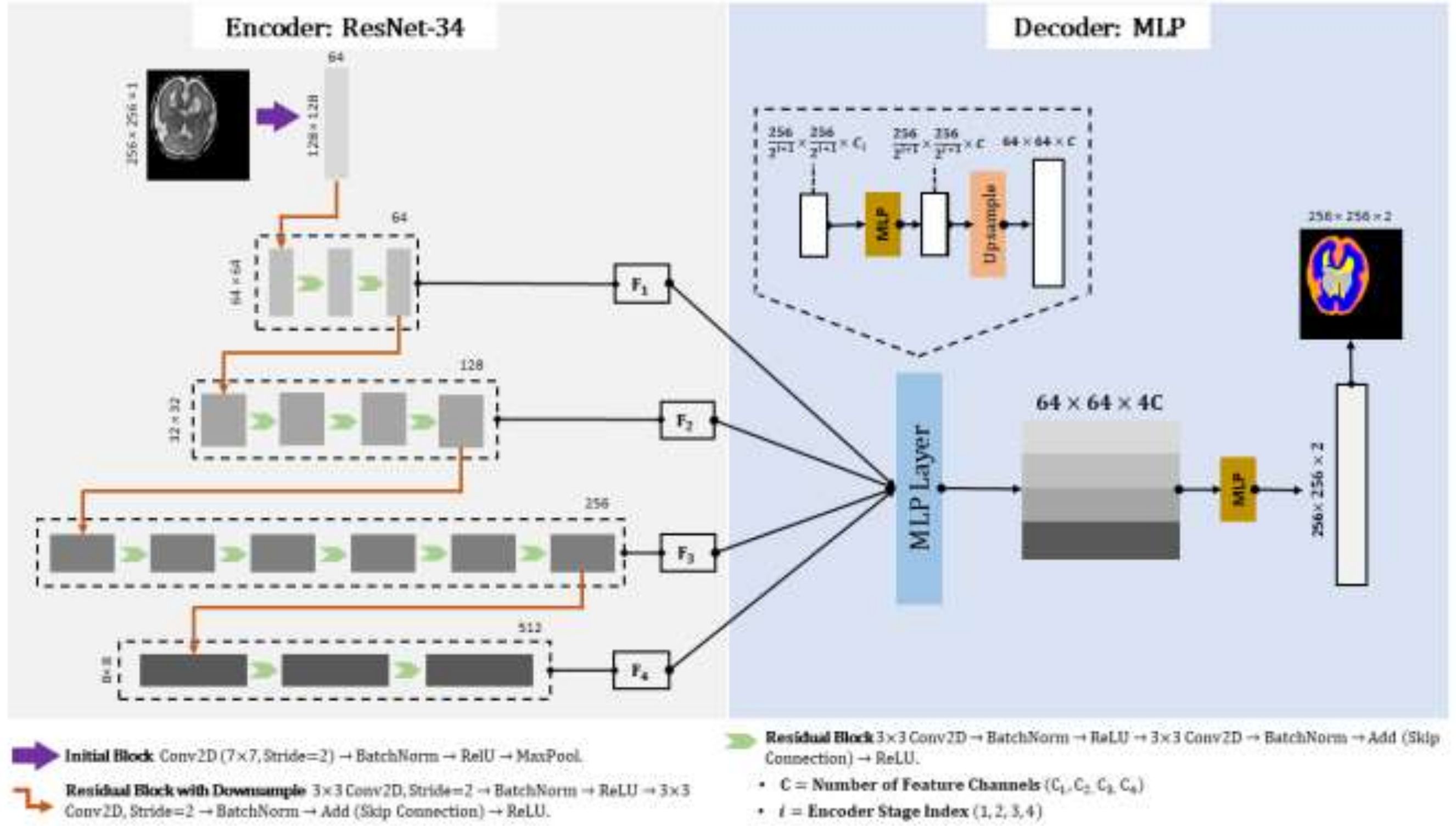

Figure 3.3: Architecture of the proposed fetal brain tissue segmentation model. The model consists of a ResNet-34 encoder and an MLP based decoder. The encoder extracts hierarchical features using residual blocks, while the decoder employs MLP modules and bilinear interpolation to fuse multi-scale features and produce high-resolution segmentation maps.

### 3.4 Encoder: ResNet-34

Our suggested model mainly depended on the ResNet-34 architecture, which is a residual network family variant (He et al., 2015). To fix the disappearing gradient issue, the deep convolutional neural network ResNet-34 adds residual connections, which makes it possible to train extremely deep networks. The encoder extracts hierarchical features from the input images, which are then passed to the decoder for segmentation. As illustrated in Figure 3.3(a), the encoder begins with a 256×256×1 greyscale image as input. Following a ReLU activation and batch normalization, the first convolutional layer applies 64 filters with a 7×7 kernel and a stride of 2. A max-pooling layer with a 3×3 kernel and a stride of 2 reduces the spatial dimensions to 128×128×64.

Then, the output from initial block goes through next four stages of residual blocks, each containing multiple residual units. The first stage (Layer 1) processes the input with 3 residual blocks, each comprising two convolutional layers with 3×3 kernels and 64 filters, producing an output of 64×64×64. The second stage (Layer 2) begins with downsampling using a 1×1 convolutional layer with a stride of 2, followed by 4 residual blocks with 128 filters, resulting in an output of 32×32×128. The third stage (Layer 3) similarly downsamples the input and processes it through 6 residual blocks with 256 filters, yielding an output of 16×16×256. The final stage (Layer 4) downsamples the input and applies 3 residual blocks with 512 filters, producing high-level feature maps of size 8×8×512.

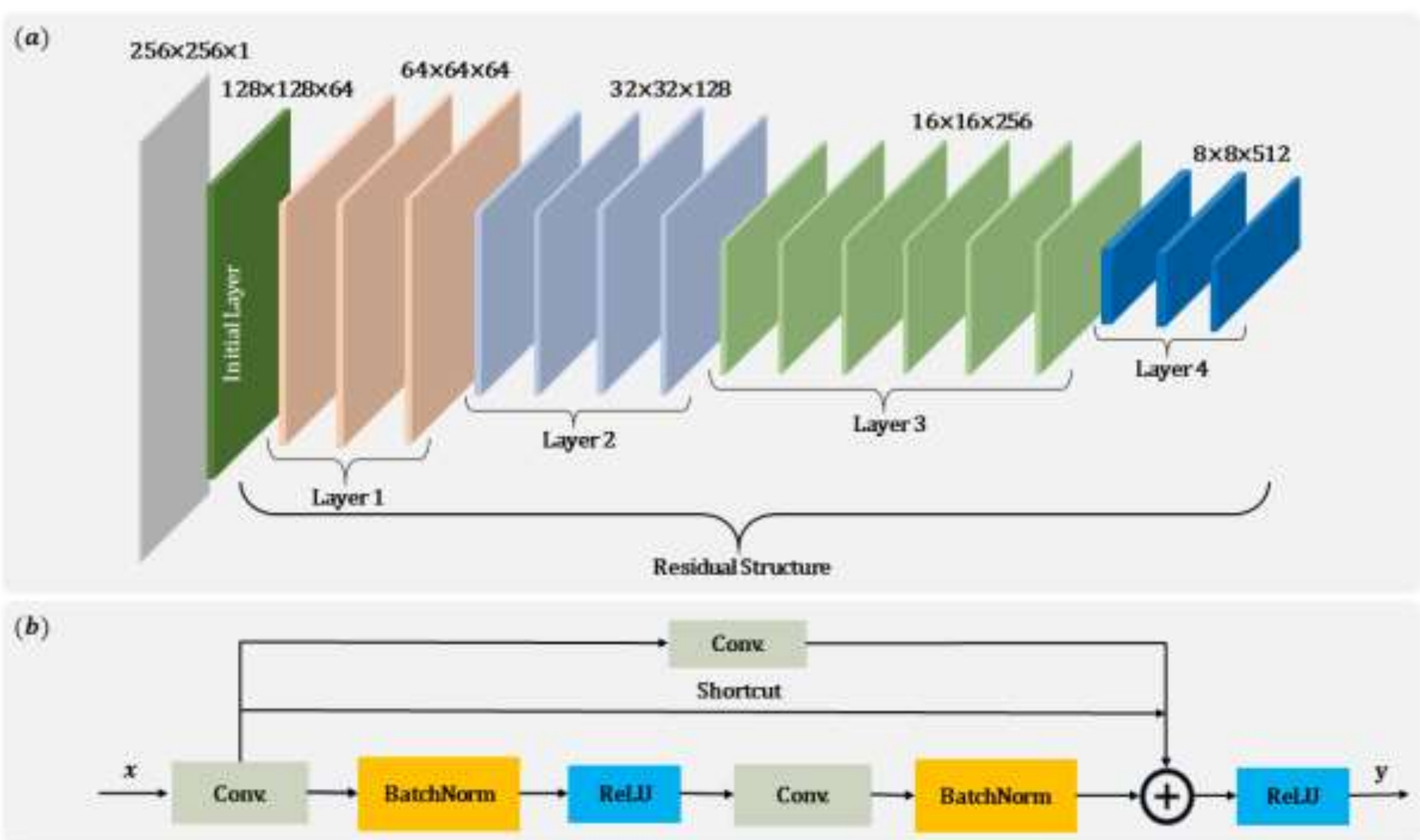

Figure 3.4: Architecture of the ResNet-34 Encoder. (a) Overview of the encoder structure, showing the input image and the hierarchical feature extraction process across four layers. The spatial dimensions and number of filters are indicated for each layer. (b) Detailed view of the residual block structure, highlighting the convolutional layers (Conv), batch normalization (BatchNorm), ReLU activation, and the skip connection (Shortcut) that enables residual learning.

As seen in Figure 3.3(b), each residual block is made up of two convolutional layers with 3 × 3 kernels, batch normalization, and ReLU activation in addition to a skip connection that adds the block's input straight to its output. This residual structure allows the network to learn residual functions, improving gradient flow and training stability. Mathematically, the output $y$ of a residual block is given by:

$$y = F(x, \{W_i\}) + x \quad 3$$

Where x is the input to the block, $F(x, \{W_i\})$ stands for the transformation applied by the convolutional layers, and $W_i$ denotes the weights of the layers. The encoder's ultimate output is a set of high-level feature maps capturing rich semantic information, which is passed to the decoder for further processing. The ResNet-34 encoder serves as a robust feature extractor, providing the decoder with high-quality feature maps for precise segmentation of fetal brain tissues.

### 3.5 Decoder: MLP-Based Decoder

In our proposed model, we employ an MLP-Based Decoder to refine the hierarchical feature maps extracted by a ResNet-34 encoder for precise fetal brain tissue segmentation. This decoder is designed to effectively aggregate multi-scale spatial and contextual information while preserving fine structural details necessary for accurate tissue delineation (Xie et al., 2021). The overall architecture of the decoder is shown in Figure 3.4, which demonstrates how feature maps are processed and fused to generate the final segmented output.

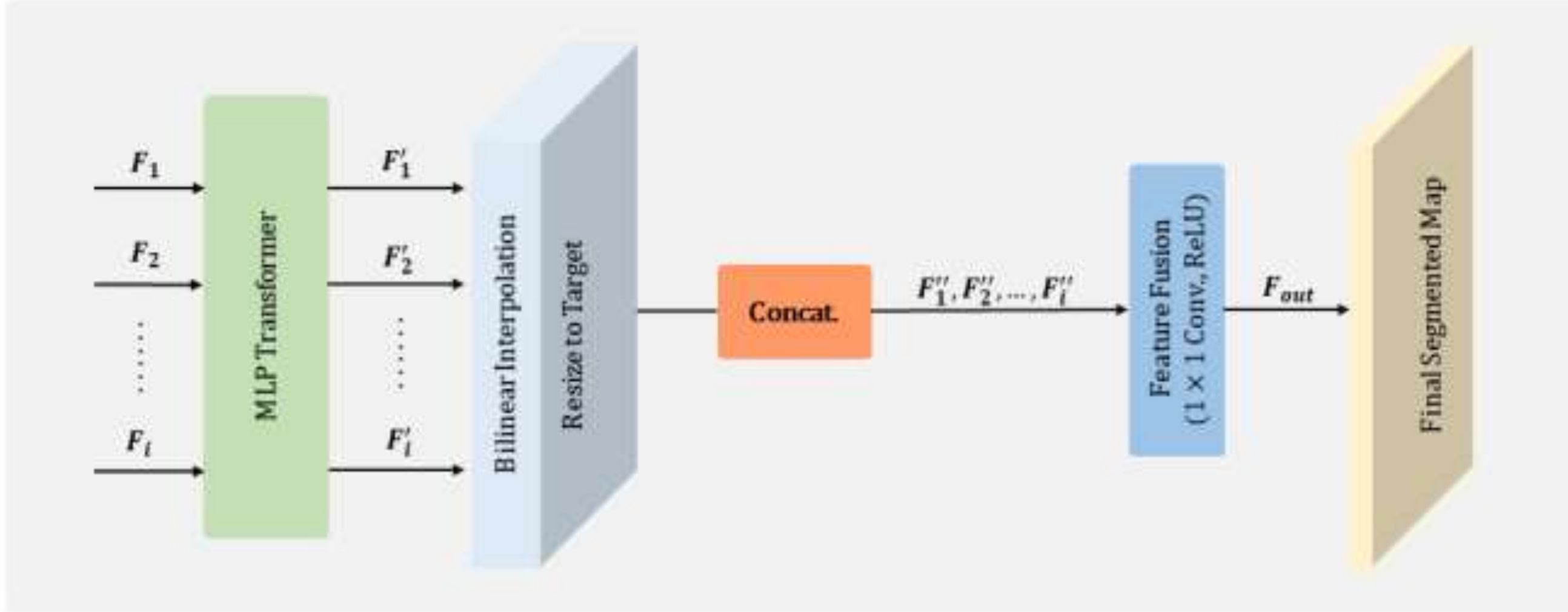

Figure 3. 5: Overview of the MLP-Based Decoder for fetal brain tissue segmentation.

ResNet-34 generates a set of multi-resolution feature maps at different encoder depths. These feature maps are first projected into a uniform latent space using Multi-Layer Perceptions to ensure consistency across different spatial scales. Each feature map $F_i$, extracted from the $i$-th ResNet-34 encoder stage, is passed through an MLP to perform a linear transformation over the channel dimension. Mathematically, this transformation is expressed as:

$$F_i' = \text{MLP}\ (F_i), \quad i \in \{1, 2, \dots, d-1\} \qquad 4$$

Where d refers to the total encoder depth, and $F_i'$ represents the MLP-projected feature. The MLP flattens the spatial structure before applying the transformation, ensuring a smooth transition between different feature resolutions.

Due to the hierarchical nature of ResNet-34, the projected feature maps $F_i'$ have varying spatial resolutions. To facilitate feature fusion, every feature map has been unsampled to a standard spatial resolution using bilinear interpolation. This resizing process ensures that all features align spatially before being merged. The upsampling operation is mathematically defined as:

$$F_i'' = \text{Interpolate}\ (F_i', \qquad \text{size} = H_t \times W_t) \qquad 5$$

Where $H_t$ and $W_t$ are the target spatial dimensions, determined by the largest feature map size in the decoder. This alignment step is essential for preserving meaningful spatial correspondence across different encoder depths. Concatenation and a convolutional fusion block are used to create a refined feature representation after all features have been unsampled to the same resolution:

$$F_{out} = \text{ReLU}\ (W * \text{Concat}\ (F_1'', F_2'', \dots, F_i'') + b) \qquad 6$$

Where convolutional kernel and bias are defined by W and b. This step ensures that all relevant multi-scale contextual information is preserved and refined for accurate segmentation.

### 3.6 Loss Function

In this study, the Cross-Entropy Loss is employed as the primary loss function to train the segmentation model (Mao et al., 2023). In segmentation tasks, cross-entropy loss is frequently utilized due to its effectiveness in handling multi-class classification problems. It measures the dissimilarity between the ground truth distribution and in every image, each image has predicted the probability distribution. Mathematically, the cross-entropy loss $\mathcal{L}_{\mathrm{CE}}$ for a single pixel is defined as:

$$\mathcal{L}_{\mathrm{CE}} = -\sum_{c=1}^{C} y_c \log(p_c) \quad 7$$

Where C is the number of classes, $y_c$ is the ground truth label for class c, and $p_c$ is the predicted probability for class c. For the entire image, the total cross-entropy loss is computed as the average loss over all pixels:

$$\mathcal{L}_{\mathrm{CE}} = -\frac{1}{N}\sum_{i=1}^{N}\sum_{c=1}^{C} y_{i,c} \log(p_{i,c}) \quad 8$$

Where N is the image's total pixel count, $y_{i,c}$ is defined as the ground truth label for pixel $i$ and class $c$, and $p_{i,c}$ is the predicted probability for pixel $i$ and class $c$. The cross-entropy loss penalizes the model for incorrect predictions which encourages it to develop accurate segmentation maps.

Cross-entropy loss was chosen over alternative losses such as dice loss and jaccard loss because it offers stable gradients and better convergence for multi-class problems with unbalanced class distributions, which is typical in fetal brain segmentation where some tissue classes may be underrepresented. Furthermore, cross-entropy directly supports pixel-wise classification, making it computationally efficient and easier to optimize. Preliminary experiments with hybrid loss formulations combining cross-entropy and dice loss did not yield consistent improvements across all tissue classes. Given the presence of class imbalance and annotation uncertainty in the FeTA dataset, cross-entropy loss demonstrated greater robustness and overall stability and was therefore selected for this study.

### 3.7 Experimental Setup

All experiments were conducted using the PyTorch framework on a high-performance workstation equipped with an Intel Core i9 (13th Generation) CPU, 64 GB RAM, and an NVIDIA GeForce RTX 4090 GPU. The software environment was managed using Anaconda with Python 3.9.21, PyTorch 2.5.0, and CUDA 12.1 to ensure reproducibility and stable GPU acceleration. For dataset preparation, a strict patient-wise splitting strategy was employed, ensuring that all 2D axial slices belonging to a given fetal brain volume were assigned to the same data split to prevent data leakage. The dataset was divided into 70% training, 10% validation, and 20% testing, and a 5-fold cross-validation scheme was applied for robust performance evaluation. Although the model operates in a 2D slice-wise manner, training on all slices from each volume allows implicit learning of anatomically consistent patterns across adjacent slices; however, no explicit inter-slice continuity constraints are imposed. No explicit data augmentation was applied during training to avoid introducing anatomically unrealistic deformations in fetal brain MRI.

The model was trained using a batch size of 16 for up to 100 epochs, with early stopping applied using a patience of 20 epochs to reduce overfitting and unnecessary computation. A Reduce-on-Plateau learning rate scheduler was employed to adaptively adjust the learning rate and facilitate stable convergence. All hyperparameters were selected based on empirical evaluation and prior experience with medical image segmentation tasks. A summary of the training hyperparameters is provided in Table 3.1.

Table 3. 1: Summary of hyperparameters used for training the fetal brain tissue segmentation model.

| Hyperparameter | Value |
|---|---|
| **Optimizer** | Adam |
| **Learning Rate** | *1e*-*4* |
| **Training Epochs** | 100 |
| **Batch Size** | 16 |
| **Early Stopping Patience** | 20 epochs |
| **Learning Rate Scheduler** | Reduce on plateau (factor=0.1, patience=5 epochs) |
| **Loss Function** | Cross-Entropy Loss |

### 3.8 Evaluation Metrics

To thoroughly analyze the performance of the proposed fetal brain tissue segmentation model, several widely used metrics were employed. These metrics provide quantitative measures of the model's accuracy, overlap, and precision in predicting the segmentation masks (Rainio et al., 2024). The following metrics were used:

Accuracy measures the ratio of correctly classified pixels to the total pixels. It is defined as:

$$\text{Accuracy} = \frac{\text{TP} + \text{TN}}{\text{TP} + \text{TN} + \text{FP} + \text{FN}} \quad 9$$

Whereas True Positive (TP) denotes positive pixels were accurately predicted, True Negative (TN) represents negative pixels were perfectly predicted, False Negative (FN) represents incorrectly predicted negative pixels, and False Positive (FP) denotes positive pixel incorrectly predicted. Although accuracy is a general metric, it is useful in providing a broad overview of segmentation performance. However, it may be less sensitive in cases of class imbalance, which is common in fetal brain tissues. Nonetheless, it serves as a baseline indicator of overall model reliability.

The overlap between the ground truth and the predicted segmentation is measured by the Dice Similarity Coefficient (DSC), also termed as dice score (Raschka, 2018). It is particularly useful for evaluating segmentation tasks with imbalanced class distributions. The dice score is defined as:

$$\mathrm{DSC} = \frac{2 \times \mathrm{TP}}{2 \times \mathrm{TP} + \mathrm{FP} + \mathrm{FN}} \qquad 10$$

DSC of 1 indicates perfect overlap, while a score of 0 indicates no overlap. DSC is particularly well-suited for medical image segmentation tasks, as it compensates for class imbalance and emphasizes overlap precision. This is especially important in fetal MRI, where certain tissues (e.g., brainstem or ventricles) may occupy small regions but require high segmentation fidelity due to their diagnostic significance.

Intersection over Union (IoU), also known as the Jaccard Index, measures the ratio of the intersection area to the union area between the predicted segmentation and the ground truth (Rainio et al., 2024). It is defined as:

$$\mathrm{IoU} = \frac{\mathrm{TP}}{\mathrm{TP} + \mathrm{FP} + \mathrm{FN}} \qquad 11$$

Like the DSC, an IoU of 1 indicates perfect overlap, while an IoU of 0 indicates no overlap. IoU provides a stricter evaluation than DSC and is especially useful for assessing segmentation boundary accuracy. This is clinically relevant in fetal brain imaging where precise boundary delineation can influence the interpretation of brain development and detection of abnormalities.

Precision is the proportion of accurately predicted positive pixels among all expected positive pixels (Rainio et al., 2024). It is defined as:

$$\text{Precision} = \frac{\text{TP}}{\text{TP} + \text{FP}} \qquad 12$$

In activities involving medical imaging, where false positives might result in incorrect diagnoses, precision is especially crucial.

These metrics were computed for each tissue class and averaged across all classes to give an in-depth assessment of the model's functionality. The use of multiple metrics ensures a robust assessment of the model's ability to accurately segment fetal brain tissues. While overall accuracy is reported for completeness, class-wise DSC, IoU, and Precision are emphasized to mitigate the effects of class imbalance inherent in fetal brain tissue segmentation.

## 4. Results and Discussions

This section provides a detailed evaluation of our proposed fetal brain tissue segmentation model, highlighting its performance and efficiency compared to existing approaches. The evaluation begins with a comparison with baseline models, including U-Net, UNet++, DeepLabV3, and DeepLabV3+, using standard metrics such as DSC, IoU, Accuracy, and Precision. A visual comparison of segmentation outputs is also provided to illustrate the capacity of the model to preserve spatial coherence and capture minute details. Next, the suggested model is benchmarked against state-of-the-art approaches in fetal brain segmentation, demonstrating its competitive performance and unique contributions. The model's computational complexity is then evaluated in terms of model parameters, training time, and inference time, and compared with the baseline models to highlight its suitability for real-world applications, specifically in resource-constrained environments. In addition, a sensitivity analysis is conducted to assess model robustness under varying contrast and motion conditions**.** Finally, an ablation study evaluates the influence of major components and configurations, such as preprocessing techniques, learning rates, and optimizers, on the model's performance. The results provide insights into the optimal design choices and validate the proposed architecture. Together, these analyses highlight how well the proposed model handles complex segmentation tasks and rapidly changing anatomy of the developing fetal brain, making it a strong candidate for clinical applications.

### 4.1 Comparison with baseline models

The proposed fetal brain tissue segmentation model was evaluated against four widely used baseline architectures such as UNet, UNet++, DeepLabV3, and DeepLabV3+ with ResNet34 encoder. The results, summarized in Table 4.1, demonstrate the superior performance of our proposed model across multiple tissue types, including white matter, lateral ventricles, gray matter, deep gray matter, cerebellum, extra-cerebral cerebrospinal fluid, and brainstem. On average, the proposed model achieves overall Accuracy of 97.37%, DSC of 90.33%, IoU of 86.93%, and Precision of 90.83%, outperforming all baseline models. Additionally, qualitative visual comparisons in Figure 4.1 help assess how well each model captures anatomical boundaries and preserves fine structural details.

Table 4. 1: Performance comparison of the proposed fetal brain tissue segmentation model against baseline architectures (UNet, UNet++, DeepLabV3, and DeepLabV3+) across multiple tissue types. Metrics include Accuracy, DSC, IoU, and Precision for white matter (WM), lateral ventricles (LV), extra-cerebrospinal fluid (eCSF), gray matter (GM), cerebellum (CB), deep gray matter (DGM), and brainstem (BS) in percentage.

| **Model** | | **Metrics** | **WM** | **LV** | **GM** | **eCSF** | **DGM** | **CB** | **BS** | **Average** |
|---|---|---|---|---|---|---|---|---|---|---|
| Encoder | Decoder | | | | | | | | | |
| **ResNet34** | UNet | Accuracy | 97.53 | 97.21 | 96.57 | 96.05 | 96.12 | 97.36 | 96.11 | 96.71 |
| | | DSC | 91.12 | 88.65 | 81.56 | 81.52 | 89.27 | 88.47 | 88.79 | 87.05 |
| | | IoU | 87.61 | 86.16 | 75.56 | 76.614 | 87.61 | 87.08 | 85.46 | 83.73 |
| | | Precision | 90.12 | 88.95 | 84.51 | 82.88 | 90.43 | 89.84 | 89.04 | 87.97 |
| **ResNet34** | UNet++ | Accuracy | 97.59 | 97.08 | 96.23 | 96.11 | 96.43 | 96.76 | 96.25 | 96.64 |
| | | DSC | 88.88 | 85.55 | 78.15 | 78.98 | 88.69 | 85.69 | 84.49 | 84.35 |
| | | IoU | 84.56 | 82.70 | 72.19 | 73.43 | 86.92 | 84.13 | 82.50 | 80.92 |
| | | Precision | 89.12 | 87.31 | 83.19 | 82.58 | 90.61 | 89.71 | 89.20 | 87.39 |
| **ResNet34** | DeepLabV3 | Accuracy | 97.85 | 97.65 | 96.78 | 96.40 | 97.17 | 96.80 | 96.53 | 97.03 |
| | | DSC | 89.47 | 85.32 | 81.08 | 81.01 | 87.71 | 89.80 | 89.36 | 86.25 |
| | | IoU | 85.89 | 82.95 | 75.44 | 75.55 | 85.89 | 88.83 | 87.61 | 83.17 |
| | | Precision | 89.21 | 87.65 | 83.56 | 84.78 | 91.02 | 90.64 | 89.52 | 88.05 |
| **ResNet34** | DeepLabV3+ | Accuracy | 97.93 | 97.64 | 96.59 | 96.20 | 97.34 | 96.86 | 96.31 | 96.98 |
| | | DSC | 91.03 | 87.70 | 80.83 | 80.48 | 90.72 | 90.59 | 88.59 | 87.13 |
| | | IoU | 87.49 | 85.19 | 74.85 | 75.47 | 89.13 | 89.20 | 86.71 | 84.01 |
| | | Precision | 90.54 | 88.90 | 83.96 | 84.81 | 91.52 | 90.69 | 89.92 | 88.62 |
| **ResNet34** | **MLP*** | **Accuracy** | **97.45** | **97.81** | **96.13** | **99.07** | **97.08** | **96.71** | **97.34** | **97.37** |
| | | **DSC** | **91.18** | **89.31** | **84.48** | **92.22** | **92.26** | **91.57** | **91.29** | **90.33** |
| | | **IoU** | **87.59** | **86.44** | **78.54** | **86.07** | **90.56** | **90.12** | **89.19** | **86.93** |
| | | **Precision** | **91.11** | **89.56** | **93.09** | **87.93** | **92.14** | **91.21** | **90.77** | **90.83** |

UNet serves as the foundational model for medical image segmentation because of its symmetric encoder-decoder architecture and skip connections, which facilitate precise localization (Weng & Zhu, 2015). However, in this study, it achieves an average DSC of 87.05 %, performing best on white matter (91.12%) and deep gray matter (89.27%). However, its performance declines significantly for gray matter (81.56%), extra-cerebrospinal fluid (81.52%), and brainstem (88.79%), which is visually apparent in slices 1 and 2, where UNet fails to separate extra-cerebrospinal fluid from nearby tissue, leading to boundary confusion and false positives. In slice 3, segmentation of the lateral ventricles is disconnected and structurally inconsistent, while slice 5 highlights under-segmentation of the brainstem, with the structure nearly omitted. These patterns align with its lower precision for gray matter (84.51%) and extra-cerebrospinal fluid (82.88%), and its IoU of just 75.56% for gray matter confirms the poor overlap with ground truth in ambiguous regions. Although the model maintains an overall accuracy of 97.71 %, this masks its inability to capture fine structural details and preserve anatomical continuity, particularly in regions affected by motion or low contrast. In contrast, our proposed model, as shown in the rightmost column, clearly delineates these challenging areas, especially in slices 3, 4, and 6, preserving shape integrity and avoiding over/under-segmentation, with consistently higher DSCs in extra-cerebrospinal fluid (84.46%), brainstem (91.29%), and deep gray matter (92.26%).

UNet++ enhances UNet by incorporating dense skip connections to refine feature fusion (Zhou et al., 2018). However, in the context of fetal brain MRI segmentation, UNet++ underperforms compared to both UNet and the proposed model, achieving a lower average DSC of 84.35% and IoU of 80.92%. This decline is particularly evident in gray matter 78.15%, extra-cerebrospinal fluid 78.98%, and brainstem 84.49% of DSC, where segmentation suffers from inconsistent boundaries and frequent misclassifications. In slice 1, UNet++ merges extra-cerebrospinal fluid and gray matter, while in slice 2, it fails to isolate the brainstem, introducing substantial overlap with surrounding tissue. Slice 4 further highlights its limitations, with incomplete segmentation of the cerebellum and fragmented gray matter, suggesting a sensitivity to subtle intensity variations. Although it slightly improves precision for deep gray matter (90.61%) and maintains acceptable accuracy (96.64% overall), the model tends to produce noisy predictions in smaller structures, a drawback of the increased sensitivity from its dense connectivity. This is seen in slices 3 and 5, where the lateral ventricles are partially misclassified, and anatomical contours appear distorted. In contrast, the proposed model consistently maintains sharper boundaries and more cohesive segmentation, especially in these complex regions, demonstrating improved generalization and spatial coherence across varying fetal slice.

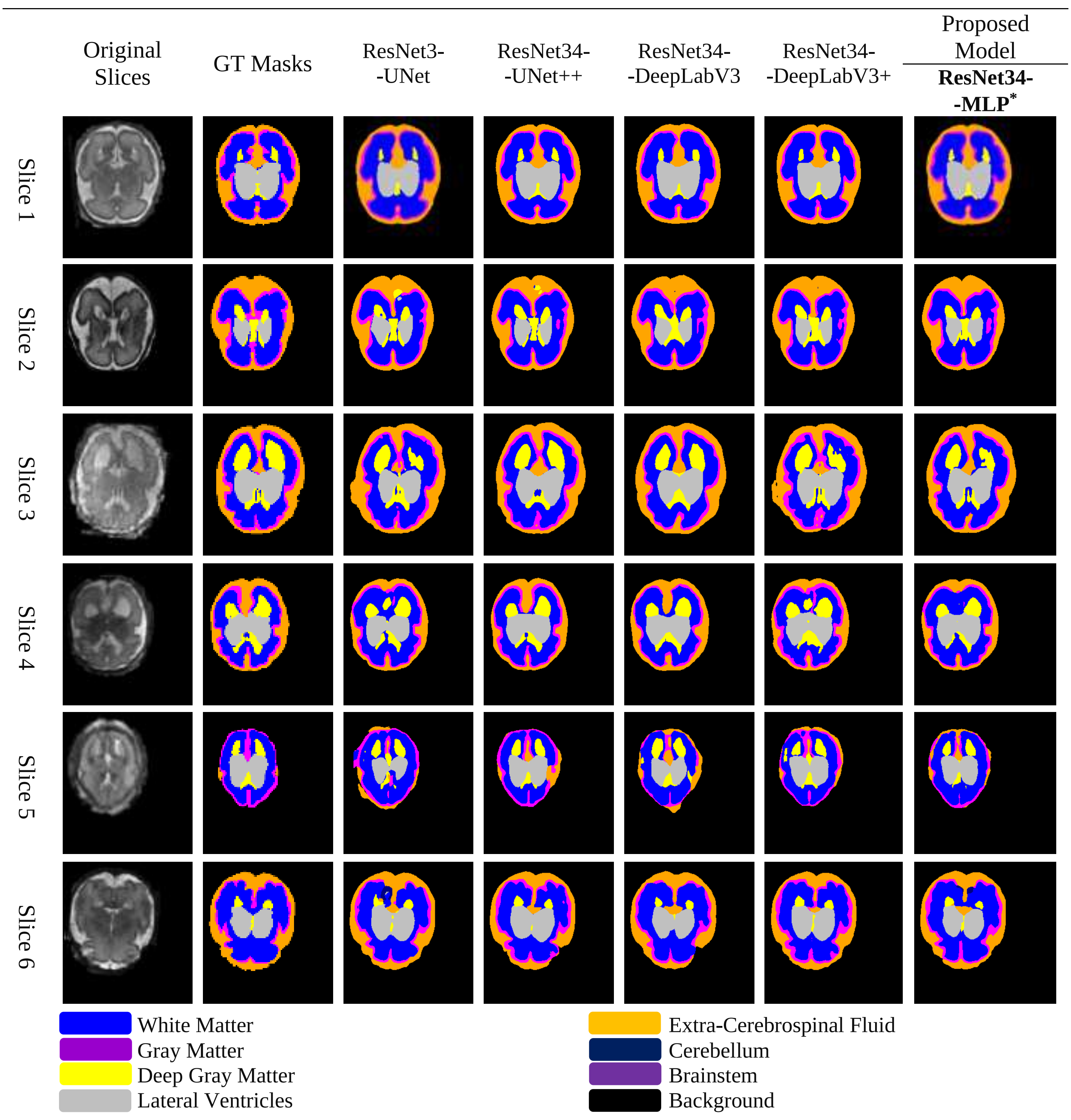

Figure 4.1: Visual comparisons of fetal brain segmentation results. The proposed model preserves fine anatomical structures, reduces over-segmentation, and improves boundary delineation compared to UNet, UNet++, DeepLabV3, and DeepLabV3+.

Notable improvements are observed in ventricles, deep grey matter, cerebellum, and brainstem, ensuring superior anatomical accuracy and structural consistency.

Atrous spatial pyramid pooling, which DeepLabV3 uses to record multi-scale contextual information, has been demonstrated to enhance medical imaging segmentation performance (L.-C. Chen et al., 2017). In our experiments, it achieves an average DSC of 86.25% and IoU of 83.17%, performing slightly better than UNet++ but still trailing behind the proposed model. The architecture shows solid performance in cerebellum (89.80% DSC) and brainstem (89.36% DSC) yet struggles with gray matter (81.08% DSC) and extra-cerebrospinal fluid (81.01% DSC), where class boundaries tend to blur. In slice 2, for example, DeepLabV3 produces visible over-segmentation of extra-cerebrospinal fluid, encroaching into surrounding cortical areas. Slice 4 highlights poor delineation of the lateral ventricles, which appear either disconnected or partially omitted. Although DeepLabV3 maintains a high accuracy of 97.03%, this masks its inconsistency in preserving fine tissue boundaries. In slice 6, the segmentation of gray matter and deep gray matter lacks spatial coherence, resulting in anatomically implausible structures. These qualitative issues, alongside lower precision in gray matter (83.56%), underscore its difficulty in handling subtle variations across fetal brain tissue. In contrast, the proposed model demonstrates more reliable region segmentation across slices, with better alignment to ground truth masks, particularly intricate areas like the ventricles, cerebellum, and brainstem.

DeepLabV3+ incorporates an Atrous Spatial Pyramid Pooling module to capture multi-scale contextual features and includes a decoder module to refine spatial details (L. C. Chen et al., 2018), which aids in segmenting larger brain regions with relatively consistent anatomical patterns. In this study, DeepLabV3+ achieved the highest overall accuracy among the baseline models (96.98%), along with an average DSC of 87.13% and an IoU of 84.01%. Notably, it performed particularly well in segmenting broad structures such as white matter (91.03% DSC), deep gray matter (90.72% DSC), and cerebellum (90.59% DSC), demonstrating robustness in regions with well-defined spatial coherence. However, its performance declined in more delicate or ambiguous regions, including gray matter (80.83% DSC), extra-cerebrospinal fluid (80.48% DSC), and brainstem (88.59% DSC). These limitations were evident in slice 1, where the boundaries between extra-cerebrospinal fluid and gray matter appeared overly merged; in slice 3, where small structures such as the ventricles were inconsistently segmented or partially omitted; and in slice 5, where the brainstem was underrepresented, leading to disconnected shapes and a loss of anatomical continuity. Although DeepLabV3+ maintained high precision in several classes, including extra-cerebrospinal fluid (84.81%) and white matter (90.54%), its reliance on global context occasionally resulted in blurred boundaries and the merging of adjacent small structures, particularly in areas with low contrast or irregular geometry. In contrast, the proposed model consistently yielded more accurate delineation of smaller structures such as the lateral ventricles and brainstem, while preserving region-specific fidelity across all six evaluated slices.

While the proposed model demonstrates superior performance across both quantitative metrics and visual evaluations, it is not without limitations. In certain low-contrast slices, particularly those affected by fetal motion or intensity inhomogeneities as seen in slices 1 and 5, the model occasionally produces soft or imprecise boundaries, especially between gray matter and extra-cerebrospinal fluid, leading to slight segmentation overlaps. In regions with subtle anatomical transitions, such as the interface between deep gray matter and surrounding tissues, the model may under-segment or blur structural distinctions when local contrast is minimal. Additionally, minor artifacts are sometimes observed in areas where the brainstem or ventricles are compressed or partially obscured, resulting in fragmented predictions or slight misalignment with the ground truth.

### 4.2 Comparison with state-of-the-art

We compare the performance of our proposed model with several state-of-the-art fetal brain segmentation techniques, as summarized in Table 4.2. A contextual transformer block integrated into an encoder-decoder architecture with hybrid dilated convolutions achieved a DSC of 83.79% on a private dataset of 80 fetal MRIs (X. Huang et al., 2023). On the FeTA 2021 and FeTA 2022 datasets, a multi scale deformable transformer with graph convolution attention achieved DSCs of 0.8666 and 0.8552 respectively (Qi et al., 2024). An Inception Residual Network with dense spatial attention modules achieved a DSC of 0.791 ± 0.18 across multiple datasets, though performance varied due to limited data diversity and modality inconsistencies (Mazher et al., 2022). A modified nnU-Net incorporating residual structures and deep supervision yielded a DSC of 0.842 on training data and 0.774 on the FeTA 2021 test set, revealing a significant performance gap between training and unseen data (Peng et al., 2023). A ResUNet with dilated convolutions, specifically designed for cerebellum segmentation in 2D ultrasound images, reported a DSC of 87.00% and precision of 90.00%, but the focus on a single brain region and use of a different imaging modality limit direct comparability (V. Singh et al., 2021).

Table 4.2: Performance comparison of the proposed model with state-of-the-art methods.

| Reference | Dataset | Modality | Classes | Method | Results |
|---|---|---|---|---|---|
| (X. Huang et al., 2023a) | 80 fetuses (private) | T2w MRI | 7 (eCSF, GM, WM, LV, CB, DGM, BS) | CoT-Block in encoder-decoder, hybrid dilated convolution | DSC: 83.79%, VS: 84.84%, HD95: 35.66 mm |
| (Qi et al., 2024) | FeTA2021 & FeTA2022 (public) | T2w MRI | 7 (eCSF, GM, WM, CB, DGM, BS, LV) | Multi-scale deformable transformers, graph convolution attention | DSC: 0.8666 (FeTA 2021), 0.8552 (FeTA 2022) |
| (Mazher et al., 2022) | Multiple datasets (private/public) | T2w MRI | 7 (eCSF, GM, WM, CB, DGM, BS, LV) | Inception Residual Network, Dense Spatial Attention Module | DSC: 0.791 ± 0.18 |
| (Peng et al., 2023) | FeTA 2021 (public) | T2w MRI | 7 (eCSF, GM, WM, VT, CB, DGM, BS) | Modified nnUNet with residual structures and deep supervision | DSC: 0.842 (training), 0.774 (FeTA 2021 test) |
| (V. Singh et al., 2021) | 734 2D fetal ultrasound images (private) | Ultrasound | Not specified (focused on cerebellum segmentation) | Residual U-Net (ResU-Net) with dilated convolutions, residual blocks | DSC: 87.00%, HD: 28.15, Recall: 86.00%, Precision: 90.00% |
| **Proposed Model** | **FetA 2021** | **T2w MRI** | **7 (eCSF, GM, WM, LV, CB, DGM, BS)** | **MLP based decoder with ResNet-34 encoder** | **Accuracy: 97.37 DSC: 90.33 IoU: 86.93 Precision: 90.83** |

*Note: classes tissue abbreviations: eCSF (extra-cerebrospinal fluid), GM (grey matter), WM (white matter), DGM (deep grey matter), CB (cerebellum), BS (brainstem).*

The proposed model, evaluated on the FeTA 2021 dataset, achieved an accuracy of 97.13%, DSC of 90.33%, IoU of 86.93%, and a precision of 90.83%, outperforming existing methods across all major metrics. Unlike several prior approaches that exhibit notable variability between training and test performance or across different datasets, our model maintains high segmentation accuracy with consistent generalization, even in anatomically complex regions. This stability, combined with competitive accuracy, underscores the model's reliability and robustness, making it a more effective solution for fetal brain tissue segmentation in T2-weighted MRI.

### 4.3 Computational Complexity

This section analyzes the computational efficiency of the proposed model in comparison with baseline architectures, including UNet, UNet++, DeepLabV3, and DeepLabV3+, all implemented with a ResNet-34 encoder. Efficiency is evaluated in terms of model parameters, training time, and inference time, which collectively indicates the scalability and practical feasibility of the model for medical imaging applications.

The quantitative comparison is summarized in Table 4.3. The proposed model demonstrates a favorable balance between segmentation performance and computational efficiency, requiring fewer parameters and shorter training and inference times than the baseline architectures. Specifically, the proposed model contains 21.87 million parameters and requires 2.03 hours for five-fold training, while achieving an inference time of 0.0211 second per image. In comparison, the baseline models require between 22.44 and 26.08 million parameters and exhibit longer training and inference times. Inference time is reported per image and reflects only the forward pass of the network, excluding preprocessing and data loading.

Table 4.3: Computational complexity comparison of the proposed model with baseline architectures.

| Model | Parameters (Millions) | Training Time (5-Fold, Hours) | Inference Time (sec/image) |
|---|---|---|---|
| **ResNet34-UNet** | 24.44 | 3.13 | 0.0433 |
| **ResNet34-UNet++** | 26.08 | 3.27 | 0.0477 |
| **ResNet34-DeepLabV3** | 26.01 | 3.22 | 0.0466 |
| **ResNet34-DeepLabV3+** | 22.44 | 2.31 | 0.0301 |
| **Proposed Model** | **21.87** | **2.03** | **0.0211** |

The reduced computational footprint of the proposed architecture highlights its suitability for practical deployment in medical imaging environments, where efficient processing is often required. Lower parameters count and faster inference enable rapid segmentation while maintaining high accuracy, which can facilitate integration into clinical workflows. Nevertheless, several challenges remain for real-world deployment. Variability in scanner types, imaging protocols, and patient demographics may influence model generalization, and performance may differ across hardware platforms. Furthermore, integration into clinical systems requires compatibility with existing software infrastructure, adherence to regulatory standards, and the development of user-friendly interfaces. Addressing these aspects will be important for translating the proposed framework from research settings to routine clinical practice.

## 4.4 Ablation and Sensitivity Analysis

To validate the design choices of the proposed fetal brain tissue segmentation framework, we conducted structured ablation experiments and sensitivity analyses. Each experiment isolates a single factor while keeping all other components fixed. In addition to evaluating preprocessing and optimization strategies, we also assessed how intrinsic image characteristics, specifically intensity contrast and motion-induced blur, influence segmentation performance.

### 4.4.1 Effect of Preprocessing

To quantify the contribution of contrast enhancement to segmentation performance, the proposed architecture was trained and evaluated with and without CLAHE under identical experimental settings. CLAHE operates by performing localized histogram equalization within small image tiles while constraining intensity amplification via a clip limit parameter. In this study, a clip limit of 2.0 and a tile grid size of 8 × 8 were adopted. The clip limit restricts excessive contrast amplification that may otherwise enhance noise, whereas the tile size controls the spatial scale at which local intensity redistribution occurs. This configuration was selected to enhance boundary delineation without introducing artificial intensity discontinuities.

As summarized in Table 4.4, the absence of CLAHE leads to a substantial decline in segmentation accuracy across all metrics.

Table 4.4: Impact of CLAHE preprocessing on the performance of the suggested fetal brain tissue segmentation model.

| Preprocessing | Accuracy (%) | DSC (%) | IoU (%) | Precision (%) |
|---|---|---|---|---|
| No CLAHE | 94.10 | 76.54 | 74.05 | 75.07 |
| CLAHE | 97.37* | 90.33* | 86.93* | 90.83* |

In particular, the DSC improves by 13.79% following contrast enhancement, accompanied by comparable gains in IoU and Precision. The magnitude of this improvement indicates that local intensity redistribution substantially strengthens boundary separability between adjacent tissue classes. This effect is especially relevant in fetal MRI, where subtle grayscale transitions between gray matter, white matter, and extra-cerebrospinal fluid often limit reliable feature extraction. These findings confirm that contrast normalization is not merely a cosmetic preprocessing step but a critical factor influencing segmentation fidelity.

### 4.4.2 Contrast-Stratified Sensitivity Analysis

To quantify the influence of intrinsic contrast variability on segmentation performance, we performed a contrast-stratified evaluation using the CLAHE-preprocessed test set. For each slice $I$, contrast was estimated as the standard deviation of pixel intensities

$$C(I) = \sigma(I) \tag{14}$$

where $\sigma(I)$ denotes intensity dispersion. Slices were ranked according to $C(I)$and partitioned into two subsets: the top 10% were designated as high-contrast slices, and the bottom 10% as low-contrast slices. The proposed model was then evaluated separately on these subsets to assess sensitivity to natural contrast variation. The quantitative results are presented in Table 4.5.

Table 4.5: Contrast-stratified performance across models.

| Model | High Contrast | Low Contrast | DSC Drop (%) |
|---|---|---|---|
| | DSC (%) | DSC (%) | |
| ResNet34-UNet | 92.15 | 81.95 | 10.2 |
| ResNet34-Unet++ | 91.26 | 77.44 | 13.82 |
| ResNet34-DeepLabV3 | 91.48 | 81.02 | 10.46 |
| ResNet34-DeepLabV3++ | 93.21 | 81.05 | 12.16 |
| **Proposed Model** | **94.47** | **86.19** | **8.28** |

All architectures exhibit performance degradation under low-contrast conditions, confirming that reduced boundary visibility negatively affects segmentation fidelity. However, the proposed MLP-based decoder achieves the highest absolute Dice scores in both subsets and demonstrates the smallest reduction (8.28%) relative to the baseline decoders. In contrast, performance drops for UNet++, DeepLabV3+, and UNet range from 10.20% to 13.82%. Because the encoder backbone is identical across models, these differences are attributed to decoder-level feature integration. The reduced sensitivity observed in the proposed architecture suggests that structured multi-scale fusion enhances boundary stability in regions with subtle grayscale transitions, thereby improving robustness under low-contrast conditions.

### 4.4.3 Motion-Stratified Sensitivity Analysis

To quantify the impact of motion-induced degradation on segmentation performance, we conducted a motion-stratified evaluation on the test set. Motion severity was approximated using an image sharpness metric derived from the variance of the Laplacian operator. For each slice $I$, the motion score was computed as

$$\mathrm{M(I)} = \mathrm{Var}(\nabla^2 \mathrm{I}) \qquad 13$$

where $\mathrm{Var}(\nabla^2 I)$represents the variance of the Laplacian response and serves as an indicator of high-frequency edge information. Lower values correspond to increased blurs and stronger motion artifacts. Slices were ranked according to $M(I)$and partitioned into two subsets: the top 10% were categorized as low-motion slices (sharpest images), while the bottom 10% were designated as high-motion slices (most blurred images). All models were evaluated separately on these subsets to assess their sensitivity to motion-induced image degradation. The quantitative results are presented in Table 4.6.

Table 4.6: Contrast-stratified performance across models.

| Model | Low Motion | High Motion | DSC Drop (%) |
|---|---|---|---|
| | DSC (%) | DSC (%) | |
| ResNet34-UNet | 94.05 | 80.05 | 14.00 |
| ResNet34-Unet++ | 93.18 | 75.52 | 17.66 |
| ResNet34-DeepLabV3 | 93.41 | 79.09 | 13.32 |
| ResNet34-DeepLabV3++ | 94.33 | 79.93 | 14.40 |
| **Proposed Model** | **96.67** | **83.99** | **12.68** |

All architectures exhibit performance degradation in high-motion slices, reflecting the adverse impact of motion-induced blur and the associated loss of high-frequency anatomical information. Nevertheless, the proposed model achieves the highest Dice scores in both subsets and demonstrates the smallest performance reduction (12.68%) relative to the baseline decoders. In comparison, the DSC drop for baseline models ranges from 13.32% to 17.66%, with UNet++ showing the largest degradation. Since all models employ the same ResNet-34 encoder, these differences are primarily attributed to the decoder architecture. The improved robustness observed in the proposed model suggests that the MLP-based multi-scale feature integration strategy helps preserve structural coherence even when edge information is partially degraded. This capability allows the model to maintain more reliable anatomical delineation under motion-corrupted imaging conditions, which are common in fetal MRI acquisitions.

### 4.4.4 Effect of Learning Rates

Optimization stability is particularly critical in fetal brain segmentation, where accurate delineation of small and low-contrast anatomical structures depends on consistent gradient updates. To examine the influence of step size on convergence behavior, the proposed model was trained using learning rates of $1 \times 10^{-3}$, $1 \times 10^{-4}$, and $1 \times 10^{-5}$, while all other training parameters were held constant. This design isolates the effect of learning rate on boundary refinement and generalization. The resulting performance metrics are summarized in Table 4.5.

Table 4. 5: Impact of learning rates on the performance of the proposed fetal brain tissue segmentation model.

| Learning Rate | Accuracy (%) | DSC (%) | IoU (%) | Precision (%) |
|---|---|---|---|---|
| $1 \times 10^{-3}$ | 96.85 | 82.81 | 79.37 | 81.61 |
| $\mathbf{1 \times 10^{-4}}$ | **97.37** | **90.33** | **86.93** | **90.83** |
| $1 \times 10^{-5}$ | 96.19 | 75.40 | 73.75 | 72.34 |

The learning rate of $1 \times 10^{-4}$ yields the highest DSC and IoU scores, indicating effective convergence and stable optimization dynamics. Increasing the rate to $1 \times 10^{-3}$results in a measurable reduction in segmentation performance, suggesting that larger parameter updates may hinder precise boundary modeling. In contrast, the lowest rate ($1 \times 10^{-5}$) produces a substantial decline in Dice and Precision, reflecting insufficient optimization progress within the fixed training schedule. These observations underscore the sensitivity of boundary-aware segmentation to the optimization scale and demonstrate that an intermediate learning rate provides the most reliable balance between convergence efficiency and structural fidelity.

### 4.4.5 Effect of Optimizers

The performance of the proposed model was indicated using two optimizers, Adam and SGD (Kingma & Ba, 2014; Ruder, 2016), and the results are summarized in Table 4.6. The Adam optimizer achieves the best performance, outperforming SGD across all metrics.

Table 4. 6: Impact of optimizers on the performance of the proposed fetal brain tissue segmentation model.

| Optimizer | Accuracy (%) | DSC (%) | IoU (%) | Precision (%) |
|---|---|---|---|---|
| **Adam** | **97.37** | **90.33** | **86.93** | **90.83** |
| SGD | 96.712 | 79.1 | 76.34 | 77.54 |

The Adam optimizer consistently outperformed SGD across all evaluation metrics. It delivered an 11.23% increase in DSC, a 10.59% increase in IoU, and a 13.29% improvement in Precision. These gains can be attributed to Adam's use of first- and second-order moment estimates to adaptively adjust learning rates during training, allowing it to handle sparse gradients and noisy data more effectively. SGD, with its fixed learning rate, showed slower convergence and was less effective in learning complex boundary information.

The use of Adam enhances both the efficiency and robustness of model training, which is crucial in time-constrained clinical environments where frequent model re-training may be required. Moreover, its superior performance in boundary-sensitive regions increases the trustworthiness of segmentations in clinical decision-making, especially when segmenting delicate fetal brain structures.

## 5. Limitation and Future Direction

While the proposed model demonstrates substantial improvements in fetal brain tissue segmentation, several limitations remain for future improvement. The most pressing concern from a clinical perspective is the limited diversity and size of the FeTA 2021 dataset, which, although well-annotated, primarily includes fetal MRIs from a gestational age range of 20 to 33 weeks. This range excludes earlier and later developmental stages, possibly restricting the model's applicability to actual fetal populations. Expanding the dataset to cover a broader gestational spectrum and include a wider variety of pathological conditions should be a primary focus, as this directly impacts clinical reliability and diagnostic utility. This study did not perform a separate quantitative evaluation for pathological versus non-pathological cases; future work will investigate pathology-specific segmentation performance using larger and more diverse datasets. Second, while the current model shows resilience against motion artifacts and low tissue contrast, it is still dependent on preprocessing techniques such as CLAHE to enhance image visibility. Although CLAHE improves segmentation in many cases, it can occasionally introduce artificial boundaries or intensity distortions. Reducing reliance on such preprocessing, or adopting learned contrast enhancement methods, could improve robustness and reduce variability in clinical deployment. Third, the model is currently designed for 2D segmentation, which limits its capacity to fully exploit volumetric MRI data. This constraint is particularly relevant for anatomically complex structures like the lateral ventricles and cerebellum, where 3D continuity is essential. Extending the proposed framework to 3D segmentation by adopting volumetric convolutions could further improve spatial coherence, but would require addressing increased memory demands, limited labeled data, and sensitivity to fetal MRI reconstruction artifacts. Finally, although the model exhibits computational efficiency, clinical deployment requires further validation across diverse hardware platforms, MRI scanner protocols, and healthcare environments. Prioritizing integration into clinical workflows, real-time performance testing, and regulatory compliance will be crucial for successful translation. Additionally, improving model interpretability and incorporating uncertainty quantification could significantly enhance trust among clinicians, making the system more transparent and actionable in diagnostic decision-making.

## 6. Conclusion

In this work, we proposed a deep learning framework for fetal brain tissue segmentation that integrates a ResNet-34 encoder with an MLP-based decoder designed for effective multi-scale feature refinement. The model was evaluated on the FeTA 2021 dataset using five-fold cross-validation and consistently outperformed baseline architectures including UNet, UNet++, DeepLabV3, and DeepLabV3+, achieving higher segmentation accuracy, DSC, IoU, and precision across all tissue classes. Notable improvements were observed in anatomically complex structures such as deep gray matter, lateral ventricles, and cerebellum. Robustness analyses further demonstrated reduced sensitivity to low tissue contrast and motion-induced image degradation compared with baseline decoders. These improvements were achieved while maintaining relatively low computational complexity, enabling efficient inference suitable for clinical environments. Although the study is limited by the dataset size and gestational age distribution of the FeTA 2021 dataset, the proposed approach provides a robust and scalable solution for automated fetal brain tissue segmentation and offers a promising direction for future work involving larger multi-center datasets and full 3D volumetric modeling in prenatal neuroimaging.

**Data Availability Statement**

The preprocessed data utilized in this study can be obtained upon request from the corresponding author.

**Conflict of Interest**

The authors affirm that there are no conflicts of interest associated with this research.